\documentclass[11pt,oneside,english, reqno]{amsart}
\usepackage{mathpazo}
\usepackage[scaled=0.92]{helvet}
\usepackage{courier}
\usepackage[T1]{fontenc}
\usepackage[a4paper]{geometry}
\usepackage{amsthm}
\usepackage{amssymb}
\usepackage{setspace}
\usepackage[authoryear]{natbib}
\usepackage[ruled,linesnumbered]{algorithm2e}
\usepackage[hidelinks]{hyperref}
\usepackage{graphicx}
\usepackage{subfigure}
\numberwithin{equation}{section}
\usepackage{lscape}  
\usepackage[utf8]{inputenc}
\usepackage{soul}
\usepackage{bm}
\usepackage{bbm}
\usepackage{caption}
\usepackage{booktabs,siunitx}
\makeatletter

\theoremstyle{plain}

\numberwithin{equation}{section}

\makeatother
\usepackage{amsmath}

\usepackage{amsfonts}
\usepackage{mathtools}
\usepackage{xcolor,color}

\usepackage{babel}

\begin{document}

\title[Forecasting realized covariances]{Forecasting realized covariances using HAR-type models}

\author{Matias Quiroz$^{\dagger}$, Laleh Tafakori$^{\ddagger}$ and Hans Manner$^{\S}$}
\thanks{$\dagger$: \textit{School of Mathematical and Physical Sciences, University of Technology Sydney, Australia}. \\ $\ddagger$: \textit{ Department of Mathematical Sciences, RMIT University, Australia}.
\\ $\S$: \textit{Department of Economics, University of Graz, Austria}}

\begin{abstract}
We investigate methods for forecasting multivariate realized covariances matrices applied to a set of 30 assets that were included in the DJ30 index at some point, including two novel methods that use existing (univariate) log of realized variance models that account for attenuation bias and time-varying parameters. We consider the implications of some modeling choices within the class of heterogeneous autoregressive models. The following are our key findings. First, modeling the logs of the marginal volatilities is strongly preferred over direct modeling of marginal volatility. Thus, our proposed model that accounts for attenuation bias (for the log-response) provides superior one-step-ahead forecasts over existing multivariate realized covariance approaches. Second, accounting for measurement errors in marginal realized variances generally improves multivariate forecasting performance, but to a lesser degree than previously found in the literature. Third, time-varying parameter models based on state-space models perform almost equally well. Fourth, statistical and economic criteria for comparing the forecasting performance lead to some differences in the models' rankings, which can partially be explained by the turbulent post-pandemic data in our out-of-sample validation dataset using sub-sample analyses. 
\end{abstract}

\maketitle

\newpage
\section{Introduction}
Accurate modeling and forecasting of volatility are pivotal in risk management, pricing derivatives, and guiding investment decisions in financial markets. The stock market's inherent association with the national economy underscores the importance of volatility, evident in major financial events like the stock market crash of 1987, the 2008 Lehman Brothers bankruptcy, and subsequent global financial crises. Volatility modeling, particularly with high-frequency data, has evolved over the years, moving from early models like generalized autoregressive conditional heteroskedasticity (GARCH) and stochastic volatility to more sophisticated approaches based on realized variance measures. Intraday volatility, exemplified by events such as the crash in 2010 and trading errors by Knight Capital in 2012, underscores regulators' need to thoroughly investigate the relationship between high-frequency trading and intraday fluctuations to mitigate potential risks. The ability to accurately forecast volatility becomes increasingly crucial in today's financial markets, particularly during instability as triggered by, e.g., Brexit and the COVID-19 pandemic.  Financial market volatility quantifies the variability in asset returns and constitutes a fundamental component in pivotal aspects of financial decision-making, including trading of volatility, asset allocation, risk management, and derivatives and asset pricing; see, e.g., \cite{caporin2017chasing}, \cite{harvey2023score} and \cite{ kikuchi2018minimum}.
Early attempts at modeling financial market volatility, dating back to the 1960s, employed temporal models such as those proposed by \cite{fama1965behavior}. These models, often based on autoregressive conditional heteroskedasticity (GARCH) frameworks \citep{engle1982autoregressive, bollerslev1986, bollerslev1994arch}, demonstrated robust in-sample parameter estimation but exhibited limitations in predicting daily squared returns (\citealp{poon2003forecasting}).

For a price process lacking jumps, volatility quantifies the magnitude of price fluctuations and is defined as the square root of integrated volatility (IV). Historically, researchers relied on low-frequency data for IV estimation, such as daily squared returns, despite the resulting high variance of such estimates. With the advent of high-frequency (intraday) data, \cite{andersen1998answering} and \cite{ andersen1998deutsche} introduced realized variance (RV) as an estimator of latent variance. This model-free metric is calculated by aggregating the squared price changes observed within each trading day from high-frequency data, offering a significant improvement in volatility estimation and prediction accuracy compared to GARCH and stochastic volatility models (\citealp{andersen2003modeling}). Furthermore, RV provides an ex-post estimate of asset return variance and is a consistent estimator of IV under the assumption of a diffusion process for the logarithmic asset price \citep{kinnebrock2008note}.

RV is advantageous for high-frequency data but faces challenges due to microstructure noise, as highlighted by studies like \cite{figlewski1997forecasting} and \cite{andersen2001distribution}. This noise, stemming from market data imperfections, introduces biases in RV estimates. To address this, researchers have developed microstructure noise-robust IV estimators to enhance accuracy in high-frequency volatility modeling. However, eliminating noise may be impractical. Alternatively, using lower-frequency data reduces the impact of noise. Still, it raises the variance of the RV estimator, posing a key challenge in achieving the optimal balance between noise reduction and variance control in high-frequency data-based volatility estimation. 
\cite{bandi2008microstructure}, \cite{patton2015good}, and \cite{liu2015does} shed light on a trade-off between the effect of the estimator and microstructure noise. They found that using 5-minute data for estimating RV is the recommendable frequency. Furthermore, another commonly used method is the realized kernel suggested by \cite{ barndorff2008designing}, which is popular for reducing the impact of microstructure noise when dealing with high-frequency data. The realized kernel approach is particularly suitable for estimating the realized covariance matrix, ensuring positive definiteness; see \cite{barndorff2011multivariate}. This is the method we rely on for realized covariance estimation in this paper, and we use it in combination with 1-minute returns due to its ability to mitigate microstructure noise. 

 The use of high-frequency measures results in more accurate variance forecasts compared to those derived from GARCH or stochastic volatility (SV) models tailored for daily returns, as demonstrated by  \cite{francq2019garch} and \cite{koopman2005forecasting}. Moreover, enhancing GARCH and SV models with RV measures derived from high-frequency data leads to enhanced model fit and forecasting accuracy. This enhancement is evidenced in studies such as \cite{engle2006multiple} and \cite{hansen2012realized} for observation-driven models, as well as \cite{dobrev2010information}, and \cite{takahashi2009estimating} for stochastic volatility models. However, an approach for forecasting RV directly has become extremely popular, namely the heterogeneous autoregressive (HAR) model by \cite{corsi2009simple}. The HAR model is widely adopted due to its simplicity and excellent predictive performance, making it a prominent choice in the literature (e.g., \citealp{busch2011role, souvcek2013realized,cubadda2017vector,liu2023trading}). HAR models can effectively replicate volatility persistence by capturing aggregated volatility across various interval sizes. Since the HAR model can be represented as a constrained autoregressive model of the order 20 model, parameter estimation using ordinary least squares (OLS) is straightforward. In the context of multivariate volatility, a realized covariance (RCov) matrix is formed from the returns of financial assets. \cite{chiriac2011modelling} expanded the univariate HAR to the multivariate domain, namely multivariate HAR, to model and predict the RCov. Like its univariate counterpart, the multivariate HAR model boasts a straightforward and easily implementable structure. An alternative approach for modeling realized covariance matrices is the Conditional Autoregressive Wishart by \cite{Golosnoy2012}.

\cite{huang2019volatility} observe that volatility risk fluctuates over time and influences asset pricing. However, despite this, models applied for realized volatility have primarily remained static and have not explored the dynamics of variance, skewness, or the potential evolution of information flow pace.
This limitation became evident during the 2008 financial crisis when volatility experienced a sudden and significant increase. Models with constant parameters proved inadequate in capturing this abrupt rise in volatility, potentially harming their forecasting accuracy. The widespread use of constant-parameter models is attributed to their simplicity in estimation and standard inference under accurate specification. However, this convenience comes at the cost of neglecting crucial structural changes in the volatility process, particularly over extended sample periods.
Early models seeking to move away from the assumption of constant parameters include the nonparametric and parametric model in \cite{ fan2003nonlinear} and \cite{cai2000efficient}, change-point ARCH models in \cite{davis2006structural}, regime-switching models in \cite{bollerslev1994arch} and \cite{ ang2002international}, and the mixture GARCH model developed by \cite{ haas2009asymmetric}. 
More recently, \cite{engle2013stock} have proposed multiplicative component structures, while \cite{amado2013modelling} explore both additive and multiplicative components. In a multivariate context, \cite{bauwens2013multivariate} and \cite{bauwens2016forecasting} introduce component structures for capturing lower-frequency movements. 

For modeling and forecasting RV using HAR models, in addition to the arguments above for time-variation in model parameters, \cite{bollerslev2016exploiting} point out that, after all, RV is only an estimator for the underlying volatility and therefore is characterized by measurement errors. Consequently, the OLS estimator suffers from an attenuation bias. This is solved by conditioning the coefficients on realized quarticity, an estimator for RV's variance, which results in the so-called HARQ model having superior model fit and forecasting performance. The multivariate extension for predicting realized covariance matrices using a variety of approaches is studied in \cite{bollerslev2018modeling}. Adapting the attenuation bias approach to the natural logarithm of univariate RV is suggested in \cite{WLWH20}. We propose to use this to model the log of the marginal RV in the multivariate realized covariance approach in \cite{bollerslev2018modeling} and show that this improves the resulting forecast. An alternative approach to incorporate time-varying parameters in the univariate setting is the state-space approach in \cite{bekierman2018forecasting}, modelling the persistence in the HAR models as a latent autoregressive process of order one.

To summarize our contributions, we compare and extend forecasting models for realized covariance. We build on the recommended approach from \cite{bollerslev2018modeling} that breaks the problem of forecasting realized covariances into separate models for variances and correlations. We extend this by comparing models for RV vs. $log RV$, models with and without controlling for attenuation bias, and state-space HAR models as in \cite{bekierman2018forecasting}. The models are compared for predicting daily covariances of 30 assets over approximately 12 years using statistical and economic measures for forecast comparison. 

This paper is set up as follows. In Section \ref{sec:methodology}, we introduce the models and present the theoretical framework. Section \ref{Sec:app} evaluates the benchmark models based on their fit and forecasting capabilities, while Section \ref{sec:conclusion} provides the conclusion.

\section{Methodology}\label{sec:methodology}
\subsection{Models for univariate realized volatility}\label{sec:univariate}
 In this work, we consider an asset with a price process, $P_{t}$, governed by a stochastic differential equation that captures both long-term trends and random fluctuations. We focus on the daily integrated variance, $I V_{t}$ calculated through the integral of squared instantaneous volatility $\sigma^{2}_{s}$ as follows
\begin{equation}
I V_{t}=\int_{t-1}^{t} \sigma_{s}^{2} d s.
\end{equation}

Assume that there are $M$ intraday returns in a trading day $t$ and denote the $j$th intra-day return
by $r_{t,j}$. Then the \emph{Realized Volatility} for day $t$ is defined as
\begin{equation}
RV_t=\sqrt{\sum_{j=1}^{M}{r^2}_{t,j}}.
\end{equation}
\emph{Realized volatility} $RV$ is inherently heterogeneous, encompassing multiple volatility components. It has been observed that volatility over extended time intervals tends to exert a greater influence on short-term volatility than the reverse. This can be understood intuitively: intraday speculators are influenced by long-term volatility since it shapes the expected trend, whereas long-term traders are less impacted by the fluctuations caused by intraday activities. Recognizing these characteristics of realized volatility, \cite{corsi2009simple}  introduced the heterogeneous autoregressive (HAR) model. This model, which is autoregressive in nature, captures the volatility dynamics by incorporating three distinct components.
The daily \emph{realized volatility} $RV_{t-1}$, the weekly mean \emph{realized volatility}
$RV_{t-1:t-5}$ and the monthly mean \emph{realized volatility} $RV_{t-1:t-20}$\footnote{More precisely,
$RV_{t-1:t-j}=\frac{1}{j}\sum_{i=1}^{j} RV_{t-i}$}. The HAR model has the following form:
\begin{equation}\label{eq:HAR}
RV_t =\beta_0+\beta_1RV_{t-1}+\beta_2RV_{t-1:t-5}+\beta_3RV_{t-1:t-20}+\epsilon_t,
\end{equation}
with $\epsilon_t$ a mean zero error term. The same model may be specified for $\log RV_t$, which we denote as the HARL model.

\cite{bollerslev2016exploiting} note that the HAR model suffers from an attenuation bias. Their key insight is that realized volatility is a noisy measurement of the integrated volatility, and that the variance of the measurement is not homogeneous in time as assumed in early work (see, e.g., \citealp{koopman2012analysis}) in the standard HAR model in \cite{corsi2009simple}, but depends on the integrated quarticity $IQ_t$. \cite{bollerslev2016exploiting} account for the heterogeneity by allowing time-varying coefficients by including a covariate $RQ_t$, which is the realized quarticity at time $t$. The authors demonstrate that accounting for time-varying coefficients is of less importance for the weekly and monthly lags, and thus propose the model
\begin{align}\label{eq:HARQ}
RV_t & = \beta_0 + (\beta_1 + \gamma RQ_{t-1}^{1/2} ) RV_{t-1} + \beta_2 RV_{t-5:t-1} + \beta_3 RV_{t-20:t-1}  + \epsilon_t, 
\end{align}
which is known as the HARQ model. Realized quarticity is defined as $RQ_t=\frac{M}{3}\sum_{i=1}^M r_{t,i}^4$.\footnote{It is important to note, however, that not only is $I V_{t-1}$ measured with uncertainty, but $R Q_{t-1}$ also acts as a noisy estimator for $I Q_{t-1}$.} The underlying concept of this model is that when the variance of the measurement error is substantial, resulting in a large $R Q_{t-1}^{1 / 2}$, the model exhibits reduced persistence for 
$\gamma<0$.  \cite{WLWH20} extend this idea to a model for the natural logarithm of $RV$, motivated by the common approach to model $\log RV_t$ instead of $RV_t$, as
\begin{align}\label{eq:HARQL}
\log RV_t & = \beta_0 + \left(\beta_1 + \gamma \frac{RQ_{t-1}^{1/2}}{RV_{t-1}} \right)\log RV_{t-1} + \beta_2 \log RV_{t-5:t-1} + \beta_3 \log RV_{t-20:t-1}  + \epsilon_t.
\end{align}
This model, termed HARQL, has been shown to perform very well empirically. 

\cite{bekierman2018forecasting} suggested a different approach to introduce a time-varying component to the autoregressive parameter by relying on the state space model
\begin{align}\label{eq:HARS}
	RV_t =\beta_0+(\beta_1+{\lambda_t})RV_{t-1}+\beta_2RV_{t-1:t-5}+\beta_3RV_{t-1:t-20}+\epsilon_t,
\end{align}
where the error term $\epsilon_t$ is assumed to be iid normally distributed. The state variable is driven by a Markovian process with Gaussian noise
\begin{equation}\label{harsstate}
\lambda_t = \phi \lambda_{t-1} + \eta_t, \hspace{0.7cm} \eta_t \sim N(0,\sigma^2_\eta).
\end{equation}
The idea behind their model is that the time-varying coefficient $\lambda_t$ may capture time-variation due to the measurement error and other variations. The model can directly be applied to $\log RV_t$ by replacing \eqref{eq:HARS} with
\begin{align}\label{eq:HARSL}
	\log RV_t =\beta_0+(\beta_1+{\lambda_t})\log RV_{t-1}+\beta_2 \log RV_{t-1:t-5}+\beta_3 \log RV_{t-1:t-20}+\epsilon_t.
\end{align}
Note that this model for $\log \left(\mathrm{RV}_{t}\right)$ instead of $\mathrm{RV}_{t}$ has the advantage that the assumption $\epsilon_{t} \sim N\left(0, \sigma_{\varepsilon}^{2}\right)$ is more likely to hold.  Forecasts for $RV_t$ use properties of the log-normal distribution; see \cite{bekierman2018forecasting}. \cite{bekierman2018forecasting} report much better forecasting performance for models based on $\log RV_t$. We term these state space models as HARS and HARSL, respectively. When predicting the realized variance (on the ordinary scale) for the log models, we use the standard approach that bias corrects $\exp(\widehat{\log RV_t})$ assuming the estimate is normally distributed.

The models above are univariate in nature, but they form the basis of the multivariate forecasting models realized covariance matrices that we introduce next.

\subsection{Models for realized covariance}\label{sec:multivariate}
Now consider an $N$-dimensional  price process $\bm P_t$ with spot covariance matrix $\Sigma(u)$ and integrated covariance matrix for day $t$
\[
\Sigma_t=\int_{t-1}^t \Sigma(u)du,
\]
which can, e.g., be estimated using the multivariate kernel estimator by \cite{barndorff2011multivariate}. We denote the corresponding estimates by $S_t$ and let $s_t$ be the (half-vectorisation of the) realized covariance  $S_t \in \mathbb{R}_{+}^{N^\star}$ of $N$ assets at time $t$, where $N^\star = N(N+1)/2$ (the number of unique elements in the covariance matrix). To extend the HAR model to a multivariate setting, \cite{chiriac2011modelling}  propose a parsimonious extension by modeling the vectorized covariance matrix, $s_t$, by
\begin{align}
\label{eq:mHAR}
s_t & = \bm \alpha_0 + \alpha_1 s_{t-1} + \alpha_2 s_{t-5:t-1} + \alpha_3 s_{t-20:t-1}  + \varepsilon_t,
\end{align}
where $\varepsilon_t$ is a mean-zero error vector, $\bm \alpha_0$ is of dimension $N^\star$ and $\alpha_1, \alpha_2, \alpha_3$ are scalar parameters. 
\cite{bollerslev2018modeling} note that this suffers from the attenuation bias in a similar way as in the univariate case and propose a vech HARQ model that directly extends the HARQ model above to the multivariate case by allowing $\alpha_1$ to be time-varying. The time variation is driven by the diagonal elements of the measurement error covariance matrix $\Pi_t$. Specifically, the multivariate HARQ model is given by
\begin{align}
\label{eq:mHARQ}
s_t & = \bm \alpha_0  + (\alpha_1\boldsymbol{\iota} + \alpha_{1Q}\pi_t) \circ s_{t-1} + \alpha_2 s_{t-5:t-1} + \alpha_3 s_{t-20:t-1}  + \varepsilon_t,
\end{align}
with the $N^\star$ dimensional vector $\pi_t=\sqrt{\mathrm{diag}(\Pi_t)}$ ($\circ$ denotes element-wise multiplication), which can be estimated straightforwardly from the data, $\boldsymbol{\iota}$ an $N^\star$ vector of ones and $\alpha_{1Q}$ a scalar parameter. 

Another approach is to model the marginal variances and the cross-correlations separately, the approach we heavily rely on in this paper, by using the following decomposition of the realized covariance matrix
\begin{align}
    \label{eq:DRD}
    S_t & = D_t R_t D_t,
\end{align}
where $D_t$ is a diagonal matrix with standard deviations and $R_t$ is the correlation matrix. \cite{oh2016high} propose the HAR-DRD model, where the individual variances are modeled via a univariate HAR model, and the correlation matrix is modeled analogously to \eqref{eq:mHAR}. We follow this approach and model the correlations parsimoniously through the following scalar HAR model
\begin{align*}
  \bm r_t-\bar{ \bm r}=\gamma_1 (\bm r_{t-1}-\bar{\bm r})+\gamma_2 (\bm r_{t-1:t-5}
  -\bar{\bm r})+\gamma_3 (\bm r_{t-1:t-20}-\bar{\bm r})+\bm \epsilon_t,
\end{align*}
with $\bar{\bm r}= \frac{1}{T}\sum_{i=1}^{T}\bm r_t$ and the components of $\bm \epsilon_t$ are iid mean zero errors. \cite{oh2016high} show that the predictions using this model formulation are valid correlation matrices if the estimates $\widehat{\gamma}_1, \widehat{\gamma}_2, \widehat{\gamma}_3 > 0$ and $\widehat{\gamma}_1 + \widehat{\gamma}_2 + \widehat{\gamma}_3 < 1$.  The forecast is then produced by separately forecasting each part and combined to a realized covariance forecast using \eqref{eq:DRD}. \cite{bollerslev2018modeling} propose the HARQ-DRD model, which uses the univariate HARQ specification \eqref{eq:HARQ} for the individual variances and the above model for realized correlations. \cite{bollerslev2018modeling} point out that it is possible to model the correlation matrix using \eqref{eq:mHARQ}, thus allowing for time-varying coefficients, but note that the heteroskedasticity in the measurement errors of the correlations tend to be somewhat limited. Therefore, constant parameters are recommended for modeling and forecasting the realized correlations. 

We propose to build on the DRD decomposition of the covariance matrix in \eqref{eq:DRD} and replace the univariate HARQ model with models that have been shown to provide superior empirical performance in the univariate case. To be specific, $RV_t$ is modeled using the HARQL model \eqref{eq:HARQL} as well as the state-space versions HARS in \eqref{eq:HARS} and HARSL in \eqref{eq:HARSL}.

\subsection{Estimation}\label{sec:estimation}
The estimation of the HAR and HARQ models, both based on $RV_t$ and $\log RV_t$, and for the correlation model is straightforwardly done by OLS. The state space model, on the other hand, requires maximum likelihood estimation using the Kalman filter. Consider, e.g., the HARSL model in \eqref{eq:HARSL}, which we can rewrite as
\begin{align}
\label{eq:log_RV_HAR_model_dlmform}
\log RV_t & = \alpha_0 + \alpha_1 \log RV_{t-1} + \alpha_2 \log RV_{t-5:t-1}  + \alpha_3  \log RV_{t-20:t-1}
 + \log RV_{t-1} \lambda_t  + \varepsilon_t, \nonumber \\
\lambda_t & = \phi \lambda_{t-1} + v_t.
\end{align}
Setting $y_t=\log RV_t$, $\alpha = (\alpha_0,\;\ \alpha_1,\;\ \alpha_2)^\top$, $$x_t = (1, \;\ \log RV_{t-1}, \;\ \log RV_{t-5:t-1}, \;\ \log RV_{t-20:t-1} )^\top$$ and $f_t=\log RV_{t-1} $, we can simplify  \eqref{eq:log_RV_HAR_model_dlmform} 
\begin{align}
\label{eq:log_RV_HAR_model_dlmform_simple}
y_t & = f_t \lambda_t + x^\top_t\beta + \varepsilon_t, \nonumber \\
\lambda_t & = \phi \lambda_{t-1} + v_t.
\end{align}
The Kalman filtering is implemented using the \texttt{dlm} package in R (see, e.g., \citealp{durbin2012time}, \citealp{petris2010r}), which estimates state space models of the form
\begin{align}
\label{eq:log_RV_HAR_model_dlmform_package}
y_t & = F_t \lambda_t  + \varepsilon_t, \nonumber \\
\lambda_t & = G_t \lambda_{t-1} + v_t.
\end{align}
Note that $F_t = f_t$ is time-varying and that $G_t = \phi$ is static.

\subsection{Forecast evaluation}\label{Sec:eval}
The models described above deliver one-step-ahead predictions $\widehat{S}_t$ for the realized covariance matrix $S_t$. To evaluate the predictions, we use the Frobenius norm
and the quasi-likelihood (Q-Like) loss function; see \cite{LRV13} for their motivation. The Frobenius norm is commonly used to measure the distance between two matrices,
\begin{align*}
  L_t^{f}= \sqrt{\mathrm{Tr}\left((S_t-\widehat{S_t})(S_t-\widehat{S_t})^\top\right)}.
\end{align*}
The Q-Like measure is based on the negative of the log-likelihood of a multivariate normal distribution and is defined as,
\begin{align*}
L_t^{q}= \log|\widehat{S_t}|+\mathrm{Tr}(\widehat{S_t}^{-1}S_t).
\end{align*}
We take the average losses over the entire out-of-sample period, and lower values are preferable.

Next, we consider the economic evaluation of the covariance predictions using some criteria suggested in the literature and nicely motivated and summarized in \cite{bollerslev2018modeling}. Their economic evaluations  are based on the use in 
the construction of global minimum variance (GMV) portfolios and portfolios designed to track the aggregate market. 
Based on the covariance forecast $\widehat{S}_t$ of the returns on the assets, to minimize the conditional volatility, the optimal portfolio allocation vector is 
\begin{align*}
    w_t=\frac{H_{t|t-1}^{-1}\boldsymbol{\iota}}{\boldsymbol{\iota}^\top H_{t|t-1}^{-1}\boldsymbol{\iota}},
\end{align*}
where $\boldsymbol{\iota}$ is a $n \times 1$ vector of ones. The variance of this portfolio should be smaller for better forecasts. 

Let $r_t^{(j)}$ represent the return on asset $j$ in day $t$. The turnover from day $t$ to day $t+1$ is given by
\begin{align*}
    TO_t=\sum_{j=1}^n |w_{t+1}^{(j)} -w_{t}^{(j)}\frac{1+r_t^{(j)}}{1+w^\top_tr_t}|.
\end{align*}
With proportional transaction costs $cTO_t$ (with $c$ being, e.g., 0, 1\% or 2\%), the portfolio excess return net of transaction costs is 
\begin{align*}
    r_{pt}=w^\top_t r_t-cTO_t.
\end{align*}
To assess how extreme the portfolio allocations are, we use the portfolio concentrations
\begin{align*}
    CO_t=(\sum_{j=1}^n w_{t}^{(j)2})^{1/2},
\end{align*}
and the total portfolio short positions,
\begin{align*}
    SP_t=\sum_{j=1}^n w_{t}^{(j)}\mathbbm{1}_{\{w_{t}^{(j)}<0\}}.
\end{align*}
Using a quadratic utility function, the economic value of the different models is determined by solving for $\Delta_\gamma$ in 
\begin{align*}
    \sum_{t=1}^T U(r_{pt}^k,\gamma) =\sum_{t=1}^T U(r_{pt}^l - \Delta_\gamma,\gamma),
\end{align*}
where the utility of the investor with risk aversion $\gamma$ is assumed to be
\begin{align*}
    U(r_{pt}^k,\gamma)= (1+r_{pt}^k)-\frac{\gamma}{2(1+\gamma)}(1+r_{pt}^k)^2.
\end{align*}
$\Delta_\gamma$  is the return that an investor with risk aversion $\gamma$ would be willing to pay to switch from model $k$ to $l$.

\section{Application}\label{Sec:app}
The application is based on high-frequency returns for a set of assets contained in the Dow Jones 30 index at some point in our sample period.\footnote{The data were obtained from EOD historical data via www.eodhd.com. The ticker symbols of the included assets can be found in Table \ref{Tab:Estimates_all_data}.} We use a multivariate realized kernel to estimate the daily return covariance matrix from intraday 1-minute returns from March 2008 until June 2024, resulting in 4051 daily observations of realized covariance matrices. We also computed the realized quarticities for the same period. Both quantities are computed using the \texttt{highfrequency} package in R \citep{highfrequency2022}. Returns are in percentage form (that is, multiplied by 100) in all calculations unless otherwise stated.

The data have some outliers, which we treat as follows. First, we consider a (daily) realized covariance matrix an outlier if any of its elements is more than 20 standard deviations away from its mean. Only 36 of 4051 observations were classified as outliers. We also found 44 outliers detected for the realized quarticities. The union between these two sets of outliers is 53 out of 4051 for the 30 assets. For these 53 days, we replace the realized covariance matrix with the covariance matrix of the previous day. Note that we cannot simply replace the outlier element with its mean because this does not guarantee that the covariance matrix remains positive definite. We treat the realized quarticities in the same way. 

The realized covariances are then modeled using the following 7 models, HARSL-DRD, HARS-DRD, multivariate HAR (M-HAR), univariate HARL, univariate HAR, HARQL, and HARQ, (only HARQL and HARQ use realized quarticities) outlined in Section \ref{sec:methodology}. 

The models use up to 20 lags as regressors, and hence we lose the first 20 observations, leaving a total of 4051 observations. Table \ref{Tab:Summary} shows summary statistics of variances, quarticities and correlations for all $n=30$ assets.

{\tiny
\begin{table}[]\caption{Summary statistics}\label{Tab:Summary}
\begin{center}
\begin{tabular}{llrrlrrlrr}
\hline
Ticker &  & \multicolumn{2}{l}{Variance}                       &  & \multicolumn{2}{l}{Quarticity}                       &  & \multicolumn{2}{l}{Correlations}                   \\ \cline{3-4} \cline{6-7} \cline{9-10} 
        &        & \multicolumn{1}{l}{Mean} & \multicolumn{1}{l}{Std} &  & \multicolumn{1}{l}{Median} & \multicolumn{1}{l}{IQR} &  & \multicolumn{1}{l}{Mean} & \multicolumn{1}{l}{Std} \\ \hline
AAPL   &  & 3.221                   & 6.139                  &  & 18.892                    & 184.197                &  & 0.316                   & 0.350                  \\
AMGN   &  & 2.607                   & 4.892                  &  & 13.994                    & 112.932                &  & 0.280                   & 0.334                  \\
AXP    &  & 4.415                   & 10.230                 &  & 19.165                    & 251.034                &  & 0.365                   & 0.336                  \\
BA     &  & 4.752                   & 11.870                 &  & 30.351                    & 333.274                &  & 0.331                   & 0.348                  \\
CAT    &  & 4.292                   & 7.888                  &  & 32.286                    & 358.000                &  & 0.356                   & 0.343                  \\
CRM    &  & 5.526                   & 12.282                 &  & 53.465                    & 542.026                &  & 0.280                   & 0.349                  \\
CSCO   &  & 2.852                   & 5.470                  &  & 12.279                    & 126.676                &  & 0.345                   & 0.326                  \\
CVX    &  & 3.053                   & 6.601                  &  & 15.228                    & 157.138                &  & 0.323                   & 0.345                  \\
DIS    &  & 2.983                   & 6.312                  &  & 12.193                    & 119.394                &  & 0.347                   & 0.337                  \\
GE     &  & 5.267                   & 14.062                 &  & 27.388                    & 294.294                &  & 0.322                   & 0.327                  \\
GS     &  & 4.316                   & 10.014                 &  & 26.222                    & 274.918                &  & 0.348                   & 0.332                  \\
HD     &  & 2.742                   & 5.177                  &  & 11.704                    & 106.322                &  & 0.340                   & 0.336                  \\
HON    &  & 2.728                   & 5.533                  &  & 10.178                    & 123.980                &  & 0.389                   & 0.328                  \\
IBM    &  & 2.033                   & 4.283                  &  & 6.016                     & 56.197                 &  & 0.364                   & 0.334                  \\
INTC   &  & 3.512                   & 5.245                  &  & 23.431                    & 246.868                &  & 0.322                   & 0.336                  \\
JNJ    &  & 1.312                   & 2.587                  &  & 3.168                     & 26.277                 &  & 0.311                   & 0.336                  \\
JPM    &  & 4.561                   & 11.325                 &  & 19.959                    & 256.341                &  & 0.366                   & 0.332                  \\
KO     &  & 1.433                   & 3.011                  &  & 2.815                     & 25.397                 &  & 0.296                   & 0.332                  \\
MCD    &  & 1.583                   & 3.675                  &  & 3.545                     & 33.104                 &  & 0.303                   & 0.339                  \\
MMM    &  & 2.242                   & 4.183                  &  & 8.625                     & 98.181                 &  & 0.381                   & 0.330                  \\
MRK    &  & 2.296                   & 4.154                  &  & 8.082                     & 89.121                 &  & 0.284                   & 0.342                  \\
MSFT   &  & 2.733                   & 4.223                  &  & 13.794                    & 129.022                &  & 0.331                   & 0.341                  \\
NKE    &  & 2.939                   & 5.829                  &  & 13.727                    & 145.529                &  & 0.322                   & 0.339                  \\
PG     &  & 1.483                   & 4.093                  &  & 2.826                     & 24.995                 &  & 0.285                   & 0.334                  \\
TRV    &  & 2.744                   & 6.768                  &  & 7.696                     & 82.316                 &  & 0.320                   & 0.331                  \\
UNH    &  & 3.515                   & 7.205                  &  & 16.004                    & 160.649                &  & 0.276                   & 0.342                  \\
V      &  & 2.869                   & 5.314                  &  & 12.332                    & 120.496                &  & 0.326                   & 0.334                  \\
VZ     &  & 1.878                   & 3.612                  &  & 5.252                     & 49.230                 &  & 0.273                   & 0.335                  \\
WMT    &  & 1.601                   & 3.697                  &  & 3.861                     & 29.268                 &  & 0.277                   & 0.334                  \\
XOM    &  & 2.752                   & 5.308                  &  & 10.751                    & 121.844                &  & 0.325                   & 0.342                  \\
\bottomrule 
\end{tabular}\\
\end{center}
\begin{flushleft}
\small{Note: The first column shows the mean and standard deviation of the realized variance (first column). The middle column shows the median and interquartile range (IQR) of the realized quarticity. The final column shows the mean across the assets of their time-averaged correlations, and the mean across the assets of the standard deviation (over time) of the correlations. The calculations are based on returns in percentage form, i.e.\ scaled by 100.}\\
\end{flushleft}
\end{table}
}

Table \ref{Tab:Estimates_all_data} reports the average parameter estimates over all 30 assets using all the available data. The table also shows the Frobenius norm and quasi-likelihood measures, which are losses indicating the quality of the fit. However, it is important to note that these are computed in-sample. The dramatically lower values for HARS indicate overfitting, which we examine later.

\begin{landscape}
{\tiny
\begin{table}\caption{In-sample results}\label{Tab:Estimates_all_data}
  \centering
\begin{tabular}{lcccccccccc}
    \toprule
{Models} &{$\overline{\alpha}_0$} & {$\overline{\alpha}_1$} & {$\overline{\alpha}_2$} & {$\overline{\alpha}_3$} &{$\overline{\phi}$} &{$\overline{\sigma}_\epsilon$}  &{$\overline{\sigma}_\eta$} & {$\overline{\alpha}_{1,q}$} & {$\overline{L}^{\mathrm{Frobenius}}$} & {$\overline{L}^{\mathrm{Q-Like}}$} \\
      \midrule     
M-HAR & 0.1593 & 0.1983 & 0.3884 & 0.2847   & - & - & - & - & 48.1023 & 42.4565 \\
 & \footnotesize{-} & \footnotesize{-} & \footnotesize{-} & \footnotesize{-}  & \footnotesize{-} & \footnotesize{-} & \footnotesize{-} & \footnotesize{-} &   & \\
HAR & 0.4184 & 0.2147 & 0.3311 & 0.3125  & - & 4.9211 & - & - & 46.654 & 40.3348 \\
 & \footnotesize{(0.2436)} & \footnotesize{(0.1098)} & \footnotesize{(0.1287)} & \footnotesize{(0.1202)}  & \footnotesize{-} & \footnotesize{(2.4239)} & \footnotesize{-} & \footnotesize{-} & & \\
 HARL  & -0.1138 & 0.2762 & 0.2182 & 0.3701 & - & 0.7609 & - & - & 46.1667  & 40.2713 \\
  & \footnotesize{(0.0443)} & \footnotesize{(0.0293)} & \footnotesize{(0.0388)} & \footnotesize{(0.0484)}  & \footnotesize{-} & \footnotesize{(0.0303)} & \footnotesize{-} & \footnotesize{-} & & \\
HARQ  & 0.2955 & 0.4245  & 0.2744 & 0.2446 & - & 4.8483 & - & -0.0006   & 46.3588  & 40.6368 \\
 & \footnotesize{(0.251)} & \footnotesize{(0.1191)} & \footnotesize{(0.1294)} & \footnotesize{(0.1034)}  & \footnotesize{-} &  \footnotesize{(2.3991)} & \footnotesize{-} & \footnotesize{(0.0004)}   & &\\
HARQL & -0.0300 & 0.5663 & 0.1939 & 0.3010 & - & 0.7421 & - & -0.0522   & 45.2730 & 40.2152 \\
 & \footnotesize{(0.0576)} & \footnotesize{(0.0774)} & \footnotesize{(0.0440)} & \footnotesize{(0.0420)}  & \footnotesize{-} & \footnotesize{(0.0280)} & \footnotesize{-} &  \footnotesize{(0.0078)}  & & \\
HARS & 0.4541 & 0.5463 & 0.1008 & 0.2588   & -0.0885 & 2.5564 & 1.0768 & - & 36.8527 & 33.7369 \\
 & \footnotesize{(0.1996)} & \footnotesize{(0.1666)} & \footnotesize{(0.0616)} & \footnotesize{(0.0990)}  & \footnotesize{(0.1549)} & \footnotesize{(1.7944)} & \footnotesize{(0.6525)} & \footnotesize{-} & & \\
HARSL & -0.0888 & 0.2656  & 0.1640 & 0.3239  & 0.9564 & 0.7399 & 0.0425 & - & 44.2440 & 39.4222 \\
 & \footnotesize{(0.0740)} & \footnotesize{(0.0275)} & \footnotesize{(0.0332)} & \footnotesize{(0.0556)}  &  \footnotesize{(0.0450)} & \footnotesize{(0.0315)} & \footnotesize{(0.0250)} & \footnotesize{-} & & \\
  \bottomrule
  \end{tabular}\\

\begin{flushleft}
\small{Note: Parameter estimates and in-sample loss measures for all models using all the available data for the 30 assets. For the multivariate HAR (M-HAR), $\alpha_0 \in \mathbb{R}^{n(n+1)/2}$ ($=465$ for $n=30$ assets), and $\overline{\alpha}_0$ denotes the average of the the $465$ estimates. The sample standard deviation of the $465$ estimates is shown in parentheses. Moreover, for M-HAR, $\alpha_1, \alpha_2, \alpha_3\in \mathbb{R}$, i.e.\ the bar notation is not needed for M-HAR (and the standard deviation of the single estimate is omitted). For the other models, all parameters are scalar-valued, with the parameters being different for each asset: the bar notation indicates averaging over the $30$ assets. The sample standard deviations of the $30$ estimates are shown in parentheses. Empty cells in the table indicate that the parameter is not available for the corresponding model. The Frobenius and Q-like measures are computed using the mean over the in-sample period.}\\
\end{flushleft}
\end{table} 
}
\end{landscape}

\subsection{Forecasting performance}\label{Sec:app_stat}
Model selection based on in-sample loss measures may suffer from overfitting.  We saw in the previous section that the HARS model achieved a dramatically lower loss compared to the other models. It is thus preferable to fit the model with a training set (in-sample) of the data and use a test set (out-of-sample) for evaluation, which we now do in a forecasting setting.

Recall that the statistical quality of the forecast is evaluated using the Frobenius norm of the predicted covariance matrix minus the true covariance matrix and the Q-Like measure based on the negative log density of a multivariate normal. For both metrics, lower values are preferable.

To evaluate the forecasting abilities of the different models, we therefore consider an out-of-sample evaluation. For each period, we consider the accuracy of the one-step-ahead forecast via a cross-validation approach with a rolling window of fixed size of 1000 observations (equal to the number of in-sample observations) that rolls forward one step at a time. The parameter estimates remain fairly stable as the rolling window moves forward, and thus we re-estimate the model only every 30th observation for computational convenience. This results in predictions for the period April 2012 until June 2024, a total of 3031 predictions for which we compute the loss functions. We additionally divide the losses of losses into two sub-samples: those that correspond to time periods with low-quarticity (defined as the smallest 50\% quarticity values) and high-quarticity (defined as the largest 50\% quarticity values), respectively. The results are shown in Table \ref{Tab:Losses_out_of_sample}. We computed the 90\% model confidence sets (MCS) for each loss and sub-sample, respectively, and marked the included models with an asterisk.  

The results show that the HARQL model generally has the best performance with the lowest loss in 3 out of 6 cases and 5 inclusions in the MCS. The second best model appears to be the HARSL with the lowest loss twice and five inclusions in the MCS. The other models perform clearly worse and are only included in the MCS for the cases that include most models. The worst performance is found for the HARS and M-HAR models with clearly higher losses and each only one inclusion in the MCS. The poor out-of-sample performance of the HARS model confirms the suspected overfitting observed in the in-sample results. Moreover, the poor performance of the HARQ model is in contrast with the findings in \cite{bollerslev2016exploiting,bollerslev2018modeling}, whereas the strong performance of the HARQL model highlights the importance of addressing attenuation bias.

{\tiny
\begin{table}[t]\caption{Statistical out-of-sample losses}\label{Tab:Losses_out_of_sample}
\begin{tabular}{llllllllll}
\hline\hline
\multicolumn{10}{c}{Sample 2008-2024}\\
\hline
       &  & \multicolumn{2}{l}{Full sample} &  & \multicolumn{2}{l}{Low-quarticity} &  & \multicolumn{2}{l}{High-quarticity} \\ \cline{3-4} \cline{6-7} \cline{9-10} 
Models &  & Frobenius          & Q-Like          &  & Frobenius         & Q-Like         &  & Frobenius         & Q-Like         \\ \hline

M-HAR  &  &    40.929         & 38.718 && 18.585 & 24.139 && 63.289* & 53.307    \\
HAR    &  &    40.275        & 36.914* && 17.378 & 24.439 && 63.188* & \textbf{49.397}*    \\
HARL   &  &    39.391         & 36.773* && 15.937 & 23.390 && 62.861* & 50.164*    \\
HARQ   &  &    40.741          & 39.998* && 17.241 & 24.261 && 64.257* & 55.745*    \\
HARQL  &  &    \textbf{38.946}* &\textbf{36.790}* && \textbf{15.527}* & 23.358 && 62.382* & 50.231*    \\
HARS   &  &    43.080           & 37.302* && 19.062 & 24.367 && 67.114 & 50.246    \\
HARSL  &  &    38.953*          & 36.922* && 15.702 & \textbf{23.274}* && \textbf{62.220}* & 50.578*    \\         
\hline
\end{tabular}\\
\begin{flushleft}
\small{Note: Statistical losses for the predicted covariance matrices for the out-of-sample period April 2012- June 2024. The in-sample periods cover the previous 4 years of data as the in-sample period using a rolling window approach. The losses are computed using the mean over the corresponding observations. Bold fonts mark the lowest loss in each column. An asterisk indicates that the model is included in the model confidence set of significance level $\alpha = 0.10$.}\\
 \end{flushleft}
\end{table} 
}


Next, we performed pairwise Diebold-Mariano tests in order to evaluate specific model extensions against their respective baseline models presented in Table \ref{Tab:DM_pairwise}. In particular, we compare models along three dimensions: $log RV$ models vs models for $RV$ in levels, Q-models against the non-attenuated counterparts of the models, and state-space models against fixed parameter versions. The results are not clear-cut, and we conclude the following: The $log$ versions outperform the models for $RV$ in the low quarticity periods. The HARQ model does not outperform its non-attenuated counterpart, and similarly, the evidence that the HARQL is better than the HARL is quite weak. Finally, time-variation based on the state-space approach is better only for the $log RV$ models, and even here, the evidence is not perfectly clear.

Our sample period includes the highly volatile Covid-19 period, which may drive the results. Therefore, in the appendix, we report the results after splitting the out-of-sample periods into the pre-Covid time 2012-2019 and the (post) Covid period 2020-2024; see Tables \ref{Tab:Losses_out_of_sample_2012-2019}-\ref{Tab:DM_pairwise_2020-2024}. For the pre-Covid subsample, the ranking of the losses is very similar to the full sample ranking, and the HARQL emerges as the best model. However, the results are more mixed for the 2020-2024 subsample, with HARQL and HARSL performing comparably well, albeit with a slight advantage for HARSL. The pairwise Diebold-Mariano tests for the pre-Covid period show more rejections than for the full-sample period and confirm the superiority of the models for $log RV$. For the 2020-2024 period the results are more mixed, but give evidence in favor of the $log RV$ models for the low-quarticity periods and the HARSL over the HARL. 

In general, we conclude that the HARQL model provides the best overall forecast performance, suggesting that modeling $log RV$ while addressing attenuation bias is the preferred approach for modeling and forecasting. Nevertheless, the similar performance of the HARSL model makes it a viable alternative as well.

{\tiny
\begin{table}\caption{Pairwise Diebold-Mariano tests} 
\label{Tab:DM_pairwise}
\begin{tabular}{llllllllll}
\multicolumn{10}{c}{Out of Sample 2012-2024}\\
\hline
       &  & \multicolumn{2}{l}{Full sample} &  & \multicolumn{2}{l}{Low-quarticity} &  & \multicolumn{2}{l}{High-quarticity} \\ \cline{3-4} \cline{6-7} \cline{9-10} 
Models &  & Frobenius          & Q-Like          &  & Frobenius         & Q-Like         &  & Frobenius         & Q-Like         \\ \hline
HARL vs. HAR
  &  &   0.979 & 0.996 && 0.000      & 0.000   &&0.993 & 0.998      \\
HARQL vs. HARQ
    &  &     0.238 & 0.159 && 0.000 & 0.000 && 0.295 & 0.159         \\
HARSL vs. HARS
  & &   0.140 & 0.376 && 0.017 & 0.000 && 0.154 & 0.452    \\
HARQ vs. HAR
   &  &     0.937 & 0.842  && 0.855 & 0.757 && 0.935  & 0.842  \\
HARQL vs. HARL
  &  &    0.781 & 0.889 && 0.000  & 0.960  && 0.822 & 0.880  \\
HARS vs. HAR
  &  &     0.851 & 0.965 &&  0.973 & 0.994 && 0.838 & 0.961   \\
HARSL vs. HARL
  &  &  0.000 & 0.999 &&  0.136 & 0.000 && 0.000 & 0.999  \\          
  \bottomrule
\end{tabular}\\
\begin{flushleft}
\small{Note: This table reports the p-values for the pairwise comparison of the different models by Diebold-Mariano tests for the out-of-sample period April 2012- June 2024.}\\ 
\end{flushleft}
\end{table} 
}

\subsection{Economic evaluation}\label{Sec:app_econ}
For evaluating the economic relevance of the competing forecasting models, we follow the methodology in \cite{bollerslev2018modeling} summarized in Section \ref{Sec:eval} for the same evaluation periods as above. Based on the different forecasts, we construct the global minimum variance portfolio (GMVP), with and without short-selling restrictions, and evaluate its performance using some measures. These measures are turnover (TO), portfolio concentration (CO), short positions (SP), mean and standard deviations of the realized portfolio returns, the Sharpe ratio, and $\Delta_{\gamma}$, which is the amount an investor is willing to pay to switch from a given model to the HARQL model. We selected this model as the basis for its convincing performance in terms of statistical losses. The Sharpe ratio and $\Delta_{\gamma}$ are computed for different transaction costs $c=0,1,2$ percent of the turnover and $\gamma=1,10$. Note that a positive $\Delta$ implies that HARQL is the superior portfolio, whereas negative values show that the corresponding other portfolio is preferred. The results are in Tables \ref{Tab:MVP1} and \ref{Tab:MVP2}.

{\tiny

\begin{table}\caption{Economic evaluation 2012-2024}\label{Tab:MVP1}
  \centering                                                                                     
\begin{tabular}{llccccccc}
    \toprule
                          & &  M-HAR & HAR & HARL & HARQ & HARQL & HARS & HARSL  \\
      \midrule
      TO & & 0.493  & 0.625  & 0.809  & 0.919  & 1.031 & 1.021  & 0.815  \\
      CO & & 0.422  & 0.413  & 0.411  & 0.429  & 0.425 & 0.425  & 0.414   \\
      SP & & -0.340 & -0.290 & -0.293 & -0.310 & -0.309  & -0.314 & -0.297 \\
      Mean ret.& & 3.605  & 6.789  & 7.295  & 6.001  & 8.046  & 8.439  & 6.433  \\
      Std. && 13.598  & 13.289 & 13.148 & 15.084 & 13.359  & 13.654 & 13.393\\
       \hline
       
     $c=0\%$ & Sharpe & 0.265  & 0.511  & 0.555  & 0.398  & 0.602  & 0.618  & 0.480 \\
       
        & $\Delta_1$ & 191.791  & 53.563  & 31.113  & 97.476 & 0.000  & -15.311 & 69.363    \\
       &  $\Delta_{10}$ & 203.295 & 50.383  & 21.222  & 185.251 & 0.158  & -0.864  & 71.125 \\

          $c=1\%$ &Sharpe & 0.181  & 0.401  & 0.411  & 0.256  & 0.422 & 0.444  & 0.339  \\
        &  $\Delta_1$ & 137.994 & 12.976 & 8.920 & 86.253 & 0.000  & -16.361  & 47.727 \\
       &  $\Delta_{10}$ &  149.728 & 9.990 & -0.952 & 174.159  & 0.158  & -1.803 & 49.511   \\
 
          $c=2\%$  &Sharpe & 0.096  & 0.291  & 0.268  & 0.114  & 0.243 & 0.270  & 0.197   \\
        & $\Delta_1$ & 84.225 & -27.552 & -13.276 & 75.060 & 0.000  & -17.411 & 26.119\\
       &  $\Delta_{10}$ & 96.312 & -30.408 & -23.132 & 163.068 & 0.157 & -2.737 & 27.891\\
\bottomrule
  \end{tabular}
  \begin{flushleft}
\small{Note: Global minimum variance portfolios for the out-of-sample period April 2012- June 2024 with the previous 4 years of data as the in-sample period using a rolling window approach. The table shows the turnover (TO), portfolio concentration (CO), and short positions (SP). The table also shows the annual average return and the annual standard deviation of the daily portfolio returns. The lower panel shows the Sharpe ratios and $\Delta_{\gamma}$  for various cost levels $c$.}\\
\end{flushleft}
\end{table} 
\label{Tab:minimum_variance_portf_28assets}
}

{\tiny
\begin{table}\caption{Economic evaluation 2012-2024 without short-selling}\label{Tab:MVP2}
  \centering                                   \begin{tabular}{llccccccc}
    \toprule
                          & &  M-HAR & HAR & HARL & HARQ & HARQL & HARS & HARSL  \\
      \midrule
      TO & & 0.335  & 0.383  & 0.516 & 0.565  & 0.650  & 0.621  & 0.518  \\
      CO & & 0.363  & 0.366  & 0.364 & 0.377  & 0.371  & 0.371  & 0.366  \\
      SP & & 0.000  & 0.000  & 0.000 & 0.000  & 0.000  & 0.000  & 0.000  \\
      Mean ret.& & 4.184  & 7.008  & 7.114 & 6.113  & 7.454  & 8.454  & 6.719  \\
      Std. && 13.855 & 13.122 & 13.148 & 14.854 & 13.201 & 13.482 & 13.312 \\
       \hline
       
     $c=0\%$ & Sharpe & 0.302  & 0.534  & 0.541 & 0.412  & 0.565  & 0.627  & 0.505   \\
       
        &  $\Delta_1$ & 143.808  & 18.720  & 14.320 & 66.729 & 0.000  & -41.402 & 32.129  \\
       &  $\Delta_{10}$ & 175.485 & 15.173  & 11.995 & 149.738 & 0.158  & -27.769  & 37.588    \\

          $c=1\%$ &Sharpe & 0.246  & 0.466  & 0.450 & 0.323  & 0.450  & 0.520  & 0.414  \\
        &  $\Delta_1$ & 112.260 & -7.967 & 0.894 & 58.239  & 0.0002  & -44.297 & 18.900  \\
       &  $\Delta_{10}$ &  144.060 & -11.403 & -1.422 & 141.323 & 0.158  & -30.602 & 24.394 \\

          $c=2\%$  &Sharpe & 0.189  & 0.398  & 0.358 & 0.234  & 0.335  & 0.412  & 0.324   \\
        & $\Delta_1$ & 80.741 & -34.655 & -12.533 & 49.778 & 0.0002  & -47.191 & 5.730 \\
       &  $\Delta_{10}$ & 112.790 & -37.982 & -14.841 & 132.908 & 0.157 & -33.434 & 11.197 \\
\bottomrule
  \end{tabular}

\begin{flushleft}
\small{Note: Global minimum variance portfolios not allowing for short-sales for the out-of-sample period April 2012- June 2024 with the previous 4 years of data as the in-sample period using a rolling window approach. The table shows the turnover (TO), portfolio concentration (CO), and short positions (SP). The table also shows the annual average return and the annual standard deviation of the daily portfolio returns. The lower panel shows the Sharpe ratios and $\Delta_{\gamma}$  for various cost levels $c$.}\\
\end{flushleft}
\end{table} 
\label{Tab:minimum_variance_portf_28NoSPassets}
}

The results for portfolios without a short-selling restriction, reported in Table \ref{Tab:MVP1}, show that different models achieve a comparably low portfolio standard deviation. The HARQ is the only model with a notably higher standard deviation. The model confidence set for the standard deviations (not reported) includes all models, which is also true for (almost) all scenarios considered below. The turnover is lowest for the M-HAR, followed by HAR, and highest for HARQL. Portfolio concentration and short positions are very similar across all models. The mean returns vary significantly over the models, which is attributed to chance, given that means are treated as unpredictable and left unmodeled. Their large variation contrasts with the results in \cite{bollerslev2018modeling} and drives the results, sometimes dominating the performance of the volatility predictions. Portfolio performance, considering Sharpe ratios and $\Delta_{\gamma}$, is best for HARS and HARQL for low transaction costs, but HAR and HARL are preferable for 2\% transaction costs. 

The results when imposing a short-selling restriction can be found in Table \ref{Tab:MVP2}. Regarding TO, M-HAR and HARS models give the more stable portfolio weights, whereas portfolio concentration is again comparable across models. The HAR, HARL, and HARQL models achieve the lowest portfolio standard deviation. As in the previous case, the variation in mean returns is noticeable and strongly drives the economic value of the competing portfolios. Again, the portfolio based on the HARS model has the highest mean returns and has the largest Sharpe ratios for all levels of transaction costs. For low transaction costs, the second best model is the HARQL, but for larger transaction costs, the basic HAR model is second-best in terms of Sharpe-ration and $\Delta_{\gamma}$.

In the appendix, the corresponding results for the 2012-2019 and 2020-2024 subsamples can be found; see Tables \ref{Tab:MVP3_2020-2024}-\ref{Tab:MVP4_NoSP_2020-2024}. For the first subsample, the HARQL performs well but is characterized by high turnover that results in large transaction costs for higher values of $c$. For the 2020-2024 period, the results are mixed, but the HARS model has the best economic performance due to significantly higher mean returns. In terms of portfolio standard deviations, it stands out that the HARQ model performs significantly worse than all other models for the more turbulent 2020-2024 period. The basic HAR model performs fairly well for both subsamples for large transaction costs due to its low turnover and low portfolio variance. 

Overall, these results are not as clear-cut as the statistical evaluation. We can conclude that the simpler M-HAR and HAR models lead to more stable portfolio weights and, hence, less turnover than the more sophisticated models. The mean returns vary unexpectedly much across models, which strongly drives the economic performance, but all models, except the HARQ, result in similar portfolio standard deviations. As global minimum variance portfolios are constructed to minimize portfolio volatility, this indicates that multiple models can result in fairly reliable predictions for the covariance matrix. The HARQL model, our preferred specification for statistical losses, performs fairly well and may be recommended if transaction costs are low. However, the basis HAR model does a good job for high transaction costs.

\section{Conclusion}\label{sec:conclusion}
We studied the problem of forecasting realized covariance matrices by extending the approach proposed by \cite{bollerslev2018modeling}, which relies on forecasting realized correlations and variances separately. Researchers face the decision of how to precisely model $RV$, and it is not clear which approach is best in the multivariate asset case, so we tried to shed light on this issue by focusing mainly on combining different approaches for univariate $RV$ forecasting while fixing the model for correlations. The evaluation of the models' forecasting performance was based on commonly used statistical and economic criteria using a dataset of 30 stocks that were part of the Dow Jones 30 index at some point during the sample period. The predictions were evaluated for the period 2012-2024 using a rolling window scheme, and as a robustness check, additionally split the evaluation period into two sub-periods (pre-Covid and Covid/post-Covid). 

For the statistical losses the results are clear cut and suggest that our proposed multivariate HARQL model for predicting realized covariances based on a univariate attenuation bias approach \citep{WLWH20} and the log-transformation of $RV$ is superior. However, the state-space model for $log RV$ also performs well, confirming the finding for the univariate case in \cite{bekierman2018forecasting}. Moreover, the results suggest that, in general, forecasting models based on log $RV$ are preferable over models for $RV$ itself and that accounting for the attenuation bias is only advantageous for log $RV$. In contrast, the performance of the $HARQ$ model in levels is disappointing compared to the results in \cite{bollerslev2016exploiting, bollerslev2018modeling}. 

The economic evaluation showed mixed and partially counterintuitive results, contrasting the clear-cut findings in \cite{bollerslev2018modeling} and in our statistical evaluation. The HARQL model performs well but is associated with high portfolio turnover, leading to large transaction costs. On the other hand, the basic HAR has fairly low turnover while still leading to low portfolio variance. The results are strongly driven by large differences in mean portfolio returns, which one would expect to be similar across models. The mixed results can partially be explained by the volatile post-pandemic sample period, arguably making the problem significantly more difficult, and the fact that we consider a set of 30 assets, in contrast to only 10 assets studied in \cite{bollerslev2018modeling}. Combining the findings of the statistical and economic evaluations, the HARQL model can be regarded as a recommendable specification.

Future research may investigate under which conditions statistical and economic evaluation of covariance forecasts disagree and whether realized covariance forecasting models are helpful for portfolio construction in larger-dimensional settings and during adverse market conditions. Additionally, research might look further into the predictions of realized correlations and whether time-varying parameter models can improve forecast performance. 

\newpage

\bibliographystyle{apalike}
\bibliography{bibliography}

\newpage

\appendix
\setcounter{table}{0}
\renewcommand{\thetable}{\Alph{section}\arabic{table}}
\section{Additional results}\label{append:results}

{\tiny
\begin{table}[h]\caption{Statistical out-of-sample losses 2012 to 2019}\label{Tab:Losses_out_of_sample_2012-2019}
\begin{tabular}{llllllllll}
\hline\hline
\multicolumn{10}{c}{Out of Sample 2012-2019}\\
\hline
       &  & \multicolumn{2}{l}{Full sample} &  & \multicolumn{2}{l}{Low-quarticity} &  & \multicolumn{2}{l}{High-quarticity} \\ \cline{3-4} \cline{6-7} \cline{9-10} 
Models &  & Frobenius          & Q-Like          &  & Frobenius         & Q-Like         &  & Frobenius         & Q-Like         \\ \hline

M-HAR  &  &    28.549 & 32.226 && 15.248 & 20.860 && 41.851 & 43.592    \\
HAR    &  &    28.151 & 31.249* && 14.662 & 21.283 && 41.639 & \textbf{41.216}*    \\
HARL   &  &    27.324 & 31.140* && 13.383 & 20.030 && 41.265 & 42.251    \\
HARQ   &  &    28.034 & 31.159* && 14.389 & 20.956 && 41.680 & 41.363*    \\
HARQL  &  &    \textbf{26.883}* & \textbf{31.103}* && \textbf{13.036}* & 20.000 && \textbf{40.730}* & 42.207    \\
HARS   &  &    29.746 & 31.378* && 16.159 & 21.052 && 43.334 & 41.704*    \\
HARSL  &  &    27.227 & 31.253* && 13.223 & \textbf{19.929}* && 41.231 & 42.578    \\
        
\hline
\end{tabular}
\begin{flushleft}
 \small{Note: Statistical losses for the predicted covariance matrices for the out-of-sample period April 2012- December 2019. The in-sample periods cover the previous 4 years of data as the in-sample period using a rolling window approach. The losses are computed using the mean over the corresponding observations. Bold fonts mark the lowest loss in each column. An asterisk indicates that the model is included in the model confidence set of significance level $\alpha = 0.10$.}\\
 \end{flushleft}
\end{table}
}
{\tiny
\begin{table}[h]\caption{Statistical out-of-sample losses 2020 to June 2024}\label{Tab:Losses_out_of_sample_2020-2024}
\begin{tabular}{llllllllll}
\hline\hline
\multicolumn{10}{c}{Out of Sample 2020-2024}\\
\hline
       &  & \multicolumn{2}{l}{Full sample} &  & \multicolumn{2}{l}{Low-quarticity} &  & \multicolumn{2}{l}{High-quarticity} \\ \cline{3-4} \cline{6-7} \cline{9-10} 
Models &  & Frobenius          & Q-Like          &  & Frobenius         & Q-Like         &  & Frobenius         & Q-Like         \\ \hline

M-HAR  &  &    62.143* & 49.843* && 29.531 & 33.170 && 94.814* & 66.546*    \\
HAR    &  &    61.051* & 46.620* && 26.307 & 32.969 && 95.858* & \textbf{60.296}*    \\
HARL   &  &    60.069* & \textbf{46.424}* && 24.485 & 32.279 && 95.718* & 60.595*    \\
HARQ   &  &    62.515* & 55.144* && 26.499 & 33.229 && 98.595* & 77.098*    \\
HARQL  &  &    59.617* & 46.534* && \textbf{23.821}* & 32.214 && 95.478* & 60.880*    \\
HARS   &  &    65.929* & 47.454* && 28.205 & 33.262 && 103.719* & 61.672*    \\
HARSL  &  &    \textbf{59.047}* & \textbf{46.424}* && 24.056* & \textbf{32.067}* && \textbf{94.101}* & 61.229*    \\
   
\hline
\end{tabular}
\begin{flushleft}
 \small{Note: Statistical losses for the predicted covariance matrices for the out-of-sample period January 2020- June 2024. The in-sample periods cover the previous 4 years of data as the in-sample period using a rolling window approach. The losses are computed using the mean over the corresponding observations. Bold fonts mark the lowest loss in each column. An asterisk indicates that the model is included in the model confidence set of significance level $\alpha = 0.10$.}\\
 \end{flushleft}
\end{table}
}

{\tiny
\begin{table}[h]\caption{Pairwise Diebold-Mariano tests} 
\label{Tab:DM_pairwise_2012-2019}
\begin{tabular}{llllllllll}
\multicolumn{10}{c}{Out of Sample 2012-2019}\\
\hline
       &  & \multicolumn{2}{l}{Full sample} &  & \multicolumn{2}{l}{Low-quarticity} &  & \multicolumn{2}{l}{High-quarticity} \\ \cline{3-4} \cline{6-7} \cline{9-10} 
Models &  & Frobenius          & Q-Like          &  & Frobenius         & Q-Like         &  & Frobenius         & Q-Like         \\ \hline
HARL vs. HAR
  &  &   0.000 & 0.999 && 0.000 & 0.000 && 0.000 & 1.000 \\
HARQL vs. HARQ
    &  &   0.000 & 0.999 && 0.000 & 0.000 && 0.000 & 1.000 \\
HARSL vs. HARS
  &  &   0.003 & 0.980 && 0.034 & 0.000 && 0.010 & 0.994 \\
HARQ vs. HAR
   &  &   0.156 & 0.871 && 0.200 & 0.000 && 0.202 & 0.917 \\
HARQL vs. HARL
  &  &   0.000 & 0.221 && 0.000 & 0.970 && 0.000 & 0.184 \\
HARS vs. HAR
  &  &   0.994 & 0.978 && 0.956 & 0.819 && 0.986 & 0.974 \\
HARSL vs. HARL
  &  &   0.475 & 1.000 && 0.369 & 0.011 && 0.563 & 1.000 \\          
\bottomrule
\end{tabular}\\
\begin{flushleft}
\small{Note: This table reports the p-values for the pairwise comparison of the different models by Diebold-Mariano tests for the out-of-sample period April 2012 - December 2019.}\\ 
\end{flushleft}
\end{table}
}

{\tiny
\begin{table}[h]\caption{Pairwise Diebold-Mariano tests 2020-2024} 
\label{Tab:DM_pairwise_2020-2024}
\begin{tabular}{llllllllll}
\multicolumn{10}{c}{Out of Sample 2020-2024}\\
\hline
       &  & \multicolumn{2}{l}{Full sample} &  & \multicolumn{2}{l}{Low-quarticity} &  & \multicolumn{2}{l}{High-quarticity} \\ \cline{3-4} \cline{6-7} \cline{9-10} 
Models &  & Frobenius          & Q-Like          &  & Frobenius         & Q-Like         &  & Frobenius         & Q-Like         \\ \hline
HARL vs. HAR
  &  &   0.989 & 0.870 && 0.000 & 0.000 && 0.996 & 0.903 \\
HARQL vs. HARQ
    &  &   0.280 & 0.158 && 0.000 & 0.000 && 0.334 & 0.158 \\
HARSL vs. HARS
  &  &   0.152 & 0.215 && 0.092 & 0.000 && 0.159 & 0.260 \\
HARQ vs. HAR
   &  &   0.940 & 0.842 && 0.982 & 0.999 && 0.936 & 0.842 \\
HARQL vs. HARL
  &  &   0.854 & 0.924 && 0.000 & 0.635 && 0.882 & 0.923 \\
HARS vs. HAR
  &  &   0.840 & 0.925 && 0.878 & 0.957 && 0.834 & 0.915 \\
HARSL vs. HARL
  &  &   0.000 & 0.963 && 0.224 & 0.000 && 0.000 & 0.978 \\          
\bottomrule
\end{tabular}\\
\begin{flushleft}
\small{Note: This table reports the p-values for the pairwise comparison of the different models by Diebold-Mariano tests for the out-of-sample period January 2020 - June 2024.}\\ 
\end{flushleft}
\end{table}
}

{\tiny
\begin{table}\caption{Economic evaluation 2012-2019}\label{Tab:MVP3_2012-2019}
  \centering                                                                                     
\begin{tabular}{llccccccc}
    \toprule
                          & &  M-HAR & HAR & HARL & HARQ & HARQL & HARS & HARSL  \\
      \midrule
      TO & & 0.448  & 0.571  & 0.797 & 0.866  & 1.058  & 0.995  & 0.809  \\
      CO & & 0.415  & 0.403  & 0.405 & 0.414  & 0.419  & 0.417  & 0.408  \\
      SP & & -0.339  & -0.287  & -0.301 & -0.303  & -0.319  & -0.317  & -0.305  \\
      Mean ret. & & 4.395  & 8.473  & 9.192 & 9.893  & 10.623  & 9.944  & 8.798  \\
      Std. & & 11.810 & 10.740 & 11.004 & 10.953 & 10.949 & 11.174 & 11.028 \\
       \hline
       
     $c=0\%$ & Sharpe Avg. & 0.372  & 0.788  & 0.835 & 0.903  & 0.970  & 0.889  & 0.797   \\
       
        &  $\Delta_1$ & 264.193  & 88.969  & 60.028 & 30.519 & 0.000  & 29.343 & 76.627  \\
       &  $\Delta_{10}$ & 299.226 & 80.992  & 62.345 & 30.835 & 0.158  & 38.436  & 79.880    \\

          $c=1\%$ & Sharpe Comp. & 0.278  & 0.666  & 0.663 & 0.724  & 0.742  & 0.680  & 0.619  \\
        &  $\Delta_1$ & 203.226 & 40.251 & 33.938 & 11.386  & 0.000  & 23.054 & 51.710  \\
       &  $\Delta_{10}$ &  238.390 & 32.406 & 36.279 & 11.840 & 0.158  & 32.279 & 54.986 \\

          $c=2\%$ & Sharpe Avg. & 0.190  & 0.534  & 0.488 & 0.524  & 0.508  & 0.463  & 0.446   \\
        & $\Delta_1$ & 142.285 & -8.408 & 7.907 & -7.749 & 0.000  & 16.794 & 26.820 \\
       &  $\Delta_{10}$ & 177.547 & -16.185 & 10.208 & -7.157 & 0.157 & 26.124 & 30.087 \\
\bottomrule
  \end{tabular}

\begin{flushleft}
\small{Note: Global minimum variance portfolios for the out-of-sample period April 2012-December 2019 with the previous 4 years of data as the in-sample period using a rolling window approach. The table shows the turnover (TO), portfolio concentration (CO), and short positions (SP). The table also shows the annual average return and the annual standard deviation of the daily portfolio returns. The lower panel shows the Sharpe ratios and $\Delta_{\gamma}$ for various cost levels $c$.}\\
\end{flushleft}

\end{table} 
}

{\tiny
\begin{table}\caption{Economic evaluation 2020-2024}\label{Tab:MVP3_2020-2024}
  \centering                                                                                     
\begin{tabular}{llccccccc}
    \toprule
                          & &  M-HAR & HAR & HARL & HARQ & HARQL & HARS & HARSL  \\
      \midrule
      TO & & 0.570  & 0.718  & 0.829 & 1.009  & 0.985  & 1.063  & 0.825  \\
      CO & & 0.433  & 0.428  & 0.421 & 0.455  & 0.434  & 0.439  & 0.423  \\
      SP & & -0.341  & -0.294  & -0.278 & -0.322  & -0.291  & -0.307  & -0.282  \\
      Mean ret. & & 2.340  & 4.093  & 4.257 & -0.228  & 3.921  & 6.030  & 2.649  \\
      Std. & & 16.216 & 16.787 & 16.179 & 20.296 & 16.703 & 17.094 & 16.688 \\
       \hline
       
     $c=0\%$ & Sharpe Avg. & 0.144  & 0.243  & 0.263 & -0.011  & 0.234  & 0.352  & 0.158   \\
       
        &  $\Delta_1$ & 67.637  & -7.145  & -18.465 & 212.268 & 0.000  & -91.883 & 56.905  \\
       &  $\Delta_{10}$ & 39.152 & -1.994  & -49.156 & 449.087 & 0.157  & -68.122  & 56.149    \\

          $c=1\%$ & Sharpe Avg. & 0.065  & 0.148  & 0.148 & -0.122  & 0.103  & 0.213  & 0.048  \\
        &  $\Delta_1$ & 26.137 & -33.761 & -34.032 & 214.661  & 0.000  & -84.000 & 40.896  \\
       &  $\Delta_{10}$ &  -1.961 & -28.366 & -64.672 & 451.477 & 0.156  & -60.129 & 40.142 \\

          $c=2\%$ & Sharpe Avg. & -0.012  & 0.052  & 0.034 & -0.233  & -0.028  & 0.075  & -0.062   \\
        & $\Delta_1$ & -15.303 & -60.377 & -49.599 & 217.083 & 0.000  & -76.116 & 24.916 \\
       &  $\Delta_{10}$ & -43.081 & -54.742 & -80.192 & 453.872 & 0.156 & -52.128 & 24.132 \\
\bottomrule
  \end{tabular}

\begin{flushleft}
\small{Note: Global minimum variance portfolios for the out-of-sample period January 2020-June 2024 with the previous 4 years of data as the in-sample period using a rolling window approach. The table shows the turnover (TO), portfolio concentration (CO), and short positions (SP). The table also shows the annual average return and the annual standard deviation of the daily portfolio returns. The lower panel shows the Sharpe ratios and $\Delta_{\gamma}$ for various cost levels $c$.}\\
\end{flushleft}

\end{table} 
}

{\tiny
\begin{table}\caption{Economic evaluation 2012-2019 without short-selling}\label{Tab:MVP4_NoSP_2012-2019}
  \centering                                                                                     
\begin{tabular}{llccccccc}
    \toprule
                          & &  M-HAR & HAR & HARL & HARQ & HARQL & HARS & HARSL  \\
      \midrule
      TO & & 0.309  & 0.351  & 0.512 & 0.538  & 0.667  & 0.609  & 0.521  \\
      CO & & 0.353  & 0.353  & 0.353 & 0.361  & 0.360  & 0.360  & 0.355  \\
      SP & & 0.000  & 0.000  & 0.000 & 0.000  & 0.000  & 0.000  & 0.000  \\
      Mean ret. & & 5.855  & 8.483  & 8.999 & 9.254  & 9.725  & 9.368  & 8.775  \\
      Std & & 5.823 & 8.702 & 9.207 & 9.592 & 10.015 & 9.728 & 8.967 \\
       \hline
       
     $c=0\%$ & Sharpe Avg. & 0.493  & 0.790  & 0.831 & 0.860  & 0.903  & 0.852  & 0.812   \\
       
        &  $\Delta_1$ & 166.744  & 51.767  & 30.575 & 19.624 & 0.000  & 15.886 & 39.888  \\
       &  $\Delta_{10}$ & 211.643 & 50.363  & 32.673 & 19.229 & 0.158  & 24.460  & 41.475    \\

          $c=1\%$ & Sharpe Avg. & 0.430  & 0.712  & 0.718 & 0.740  & 0.754  & 0.720  & 0.696  \\
        &  $\Delta_1$ & 130.932 & 20.133 & 15.030 & 6.688  & 0.000  & 10.115 & 25.276  \\
       &  $\Delta_{10}$ &  175.918 & 18.830 & 17.168 & 6.382 & 0.158  & 18.752 & 26.886 \\

          $c=2\%$ & Sharpe Avg. & 0.368  & 0.634  & 0.605 & 0.621  & 0.606  & 0.587  & 0.581   \\
        & $\Delta_1$ & 95.147 & -11.440 & -0.453 & -6.248 & 0.000  & 4.341 & 10.724 \\
       &  $\Delta_{10}$ & 140.190 & -12.705 & 1.660 & -6.464 & 0.157 & 13.044 & 12.296 \\
\bottomrule
  \end{tabular}

\begin{flushleft}
\small{Note: Global minimum variance portfolios not allowing for short-sales for the out-of-sample period April 2012- December 2019 with the previous 4 years of data as the in-sample period using a rolling window approach. The table shows the turnover (TO), portfolio concentration (CO), and short positions (SP). The table also shows the annual average return and the annual standard deviation of the daily portfolio returns. The lower panel shows the Sharpe ratios and $\Delta_{\gamma}$ for various cost levels $c$.}\\
\end{flushleft}
\end{table} 
}

{\tiny
\begin{table}\caption{Economic evaluation 2020-2024 without short-selling}\label{Tab:MVP4_NoSP_2020-2024}
  \centering                                                                                     
\begin{tabular}{llccccccc}
    \toprule
                          & &  M-HAR & HAR & HARL & HARQ & HARQL & HARS & HARSL  \\
      \midrule
      TO & & 0.377  & 0.436  & 0.521 & 0.611  & 0.619  & 0.639  & 0.511  \\
      CO & & 0.379  & 0.385  & 0.381 & 0.404  & 0.387  & 0.390  & 0.383  \\
      SP & & 0.000  & 0.000  & 0.000 & 0.000  & 0.000  & 0.000  & 0.000  \\
      Mean ret. & & 1.509  & 4.648  & 4.097 & 1.086  & 3.819  & 6.991  & 3.428  \\
      Std. & & 16.714 & 16.438 & 16.391 & 20.018 & 16.561 & 16.933 & 16.762 \\
       \hline
       
     $c=0\%$ & Sharpe Avg. & 0.090  & 0.282  & 0.249 & 0.054  & 0.230  & 0.412  & 0.204   \\
       
        &  $\Delta_1$ & 67.637  & -7.145  & -18.465 & 212.268 & 0.000  & -91.883 & 56.905  \\
       &  $\Delta_{10}$ & 39.152 & -1.994  & -49.156 & 449.087 & 0.157  & -68.122  & 56.149    \\

          $c=1\%$ & Sharpe Avg. & 0.040  & 0.223  & 0.179 & -0.013  & 0.147  & 0.328  & 0.136  \\
        &  $\Delta_1$ & 26.137 & -33.761 & -34.032 & 214.661  & 0.000  & -84.000 & 40.896  \\
       &  $\Delta_{10}$ &  -1.961 & -28.366 & -64.672 & 451.477 & 0.156  & -60.129 & 40.142 \\

          $c=2\%$ & Sharpe Avg. & -0.010  & 0.164  & 0.108 & -0.082  & 0.063  & 0.244  & 0.068   \\
        & $\Delta_1$ & -15.303 & -60.377 & -49.599 & 217.083 & 0.000  & -76.116 & 24.916 \\
       &  $\Delta_{10}$ & -43.081 & -54.742 & -80.192 & 453.872 & 0.156 & -52.128 & 24.132 \\
\bottomrule
  \end{tabular}

\begin{flushleft}
\small{Note: Global minimum variance portfolios not allowing for short-sales for the out-of-sample period January 2020- June 2024 with the previous 4 years of data as the in-sample period using a rolling window approach. The table shows the turnover (TO), portfolio concentration (CO), and short positions (SP). The table also shows the annual average return and the annual standard deviation of the daily portfolio returns. The lower panel shows the Sharpe ratios and $\Delta_{\gamma}$ for various cost levels $c$.}\\
\end{flushleft}
\end{table} 
}

\end{document}